\documentclass[10pt,a4paper]{article}
\usepackage[sc]{mathpazo}
\usepackage{microtype}
\usepackage[english]{babel}
\usepackage[margin=0.85in,top=32mm,columnsep=20pt]{geometry}
\usepackage[hang,small,labelfont=bf,up,textfont=it,up]{caption}
\usepackage{booktabs}
\usepackage{lettrine} 
\usepackage{enumitem}  
\setlist[itemize]{noitemsep} 
\usepackage{abstract} 
\usepackage{titlesec}  
\usepackage{fancyhdr}  
\usepackage{titling} 
\usepackage{hyperref}  
\usepackage{amsmath}
\usepackage{scalerel,amssymb}
\usepackage [english]{babel}
\usepackage [autostyle, english = american]{csquotes}
\usepackage{amssymb}
\usepackage{amsmath}

 \usepackage{authblk}
\usepackage{xcolor}
\usepackage{tikz}

\usepackage{footnote}
\usepackage[left]{lineno}
\usepackage{fancyhdr}
\usepackage{geometry}
\pretitle{\begin{center}\Huge\bfseries}  
\posttitle{\end{center}} 
\title{Can Gibbons-Hawking Radiation and Inflation Arise Due to Spacetime Quanta?}  

\author[*]{Naouel Boulkaboul}

\affil[*]{Independent Researcher}
\date{}
\renewcommand{\maketitlehookd}{
In this study, we provide an alternatively reformulated interpretation of Gibbons-Hawking radiation as well as inflation. By using a spacetime quantization procedure, proposed recently by L.C. C\'eleri et al., in anti-de Sitter space we show that Gibbons-Hawking radiation is an intrinsic property of the concerned space, that arises due to the existence of a scalar field whose quanta  ``carry"  a length $l$ (i.e. the radius of the hyperboloid curvature). Furthermore, within the context of Tsallis q-framework, we propose an inflationary model that depends on the non-extensive parameter $q$. The main source of such an inflation is the same scalar field mentioned before. Being constrained by the observational data, the q-parameter along with the rest of  the model's parameters has been used to estimate the time at which inflation ends as well as the reheating temperature. The latter is found to be related to Gibbons-Hawking temperature. Thus, the present model offers an alternative perspective regarding the nature of the cosmic background radiation (CMB).\\ 
}
\begin{document}

\maketitle
\thispagestyle{fancy}
\section{INTRODUCTION}
Black hole radiation and inflation are ones of the most intriguing aspects in cosmology. The first one being 
theorized by Stephen
Hawking \cite{1}, has been attracted a considerable interest $\cite{2}-\cite{7}$. Along with Unruh radiation that is ``felt" by an accelerated observer$\cite{8}-\cite{11}$, Hawking radiation was originally derived based on Bogoliubov’s method \cite{12}, ever since many approaches $\cite{13}-\cite{17}$ including the complex path (or Hamilton-Jacobi) method \cite{18}$-$\cite{19} have been developed to  derive such a radiation. Hawking radiation is described as a tunneling effect of particles across the black hole's horizon \cite{12x}. In fact, such a thermal radiation is related to any geometrical background that possesses a horizon (i.e. the cosmological horizon of a black hole, de Sitter space and even the one of an accelerated observer). \\\\
The hypothesis of inflation, on the other hand, has been postulated by Alan Guth $\cite{20}-\cite{21}$. His model relies on the assumption that the very early Universe has gone through a period of accelerated expansion
that preceded the standard radiation-dominated era.  Such
a period of accelerated expansion offers a
physical explanation of the cosmology's biggest puzzles that the standard cosmological scenario cannot explain. Inflation drives any initially curved spacetime towards the spatial flatness observed today, hence answering the question: ``Why would the universe be perfectly spatially flat?", it brings together all causally disconnected regions and extends the causal
horizon beyond the present Hubble length, in such a away to answer the question: ``Why would the universe have the same temperature everywhere?" and it also brings a satisfactory solution to the magnetic monopoles problem, answering the question: ``Why are there no leftover high-energy relics?"
To date, many simple as well as complex inflationary models have been proposed $\cite{22}-\cite{29}$.  In their simplest picture, inflationary models are based on a scalar field (inflaton field) with a fine-tuning potential, i.e., a flat potential. The latter is constrained by slow roll conditions, i.e., in order to trigger inflation the scalar field's potential energy must dominate over its kinetic and gradient energy that can prevent its starting. This fine-tuning requirement puts the viability of such inflationary models into question.\\\\To contribute to the ongoing literature on the two topics, firstly we address the Gibbons-Hawking radiation $\cite{30}-\cite{31}$ in anti-de Sitter space by incorporating a quantized spacetime. We want to stress, however, that our contribution is to show that this radiation has nothing to do with the tunnelling particle but it is due to the space curved geometry, precisely its quantum nature. Secondly we propose, inspired by the Tsallis q-formalism $\cite{32}-\cite{34}$, a new inflationary model, with a minimum of fine-tuning, that we will call the q-inlfation. The latter, which is devoid of singularities, is triggered implicitly by the spacetime quanta but invokes the contribution of the ``cosmological constant" carried by the quanta, explicitly. We believe in the importance of the present work because it kills two birds (Gibbons-Hawking radiation and inflation) with one stone (spacetime quantization). Throughout the paper, we consider the metric signature $+---$. Moreover, the units are chosen with $ c=\hbar=1$.\\\\
\section{THE ORIGIN OF GIBBONS-HAWKING RADIATION}
\begin{figure}
\begin{center}
\includegraphics[scale=0.25]{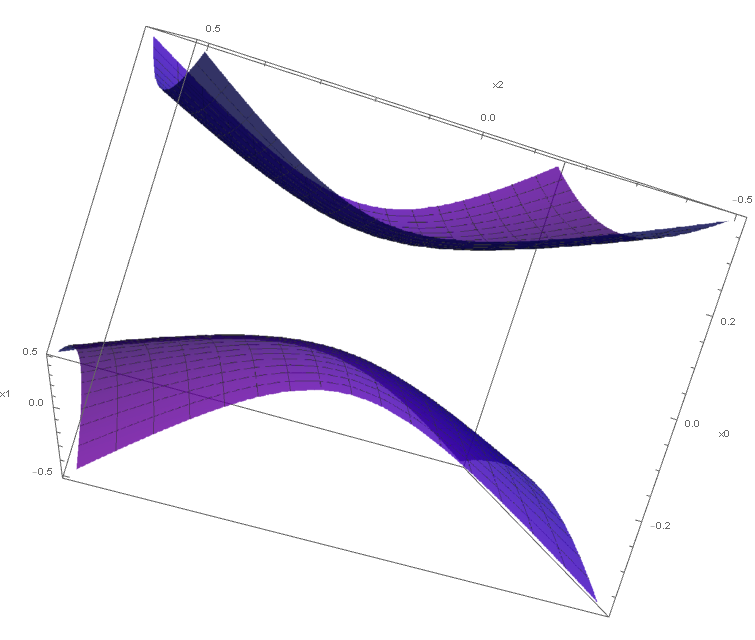}
\caption{Parameterization of two-sheet hyperboloid in the coordinates $\ref{static coord1}-\ref{static coord3}$. The ${x}_3$ and ${x}_4$ axes are suppressed.}
\label{figure:two sheet hyperboloid}
\end{center}
\end{figure}

In this section we derive the well-known Gibbons-Hawking temperature $\cite{30}-\cite{31}$ by means of a spacetime quantization framework firstly introduced by L.C. C\'eleri et al. \cite{35}$-$\cite{36}. Their study, in which they show that Unruh effect can be obtained without changing the reference frame, focuses on deriving Unruh radiation in the accelerated fields scenario. For our purpose however, let us consider the two-sheet hyperboloid (see Fig.\ref{figure:two sheet hyperboloid}) governed by the parameterization
\begin{equation}
\label{static coord1}
x_0=\sqrt{l ^2+r^2}cosh(t/l), 
\end{equation}
\begin{equation}
\label{static coord2}
x_1=\sqrt{l ^2+r^2}sinh(t/l),
\end{equation}
\begin{equation}
\label{static coord3}
x_i=rz_i, \qquad        2\leq i \leq 4.
\end{equation}
where $z_i$ gives the standard embedding of the 2-sphere in $\textbf{R}^{3}$. Embedding this in the five-dimensional Minkowski metric
\begin{equation}
\label{minkowski metric}
ds^2_M=dx_0^2-\sum\limits_{\mu=1}^{4} dx_\mu dx^\mu,
\end{equation}
yields
\begin{equation}
\label{de sitter metric}
ds^2=-\big (1+\frac{r^2}{l^2}\big )dt^2-\frac{1}{1+\frac{r^2}{l^2}}dr^2-r^2[d\theta^2+sin^2\theta d\phi^2],
\end{equation}
One may notice the strange signature $(- - - -)$ of the metric, one common way to reproduce the metric with the appropriate Lorentzian signature, is making use of the Wick rotation $t \to it$. For our convenience however we will use the real Wick rotation defined in \cite{37} (and references therein), proven to be more adequate especially for curved spacetimes. Hence, we set the above line element as
\begin{equation}
\label{de sitter metric}
ds^2=\eta_{00}dt^2+\eta_{\mu\nu}dX^\mu dX^\nu, \qquad \mu, \nu=1, 2, 3
\end{equation}
which means ``singling out" a proper time $t$, where $\eta_{00} =-\big (1+{r^2}/{l^2}\big )$, $X^1=r, X^2=\theta, X^3=\phi$ and the corresponding components of the metric tensor $\eta_{\mu \nu}$ are $\eta_{11}=-1/({1+{r^2}/{l^2}}),\eta_{22}=-r^2, \eta_{33}=-r^2sin^2\theta$.\\
Anti-de Sitter metric with Lorentzian signature is then recovered under mapping the 00-component of the metric as $\eta_{00} \mapsto -\eta_{00}$ while keeping $\eta_{\mu \nu}$ unchanged, thus we get
\begin{equation}
\label{de sitter metric}
ds_{AdS}^2=\big (1+\frac{r^2}{l^2}\big )dt^2-\frac{1}{1+\frac{r^2}{l^2}}dr^2-r^2[d\theta^2+sin^2\theta d\phi^2],
\end{equation}
 Note that, unlike de Sitter metric, this metric doesn't possess a cosmological horizon.
In order to quantize spacetime we follow the formulation of accelerated quantum field theory proposed by L.C. C\'eleri et al. $\cite{36}$. This can be achieved starting from the four-dimensional two-sheet hyperboloid's equation
\begin{equation}
\label{hyperboloid}
x^2_0-\sum\limits_{\mu=1}^{4} x_\mu x^\mu=l^2,
\end{equation}
where $x_0....x_4$ are the Cartesian coordinates in Minkowski space in which Anti-de Sitter space is embedded, and $l$ is a nonzero constant with dimensions of length (the radius of the hyperboloid's curvature).\\\\
Now we can introduce the time and position operators given by
\begin{equation}
\label{operator}
x_0=i\frac{\partial}{\partial \omega}=i\partial_\omega, \qquad x_\mu=i\frac{\partial}{\partial{ k^\mu}}=i\partial_{k^\mu}, 
\end{equation}
Making use of Eq. \ref{hyperboloid}, a wave equation similar to the Klein-Gordon equation can be obtained\\
\begin{equation}
\label{equ}
\bigg [\partial^2_\omega - \partial^2_{{{k^\mu}}}+l^2 \bigg ] {\varphi}({\omega, k^\mu})=0, 
\end{equation}
with $(\omega, k^\mu)$ being the five momentum.
We can now construct the Lagrange density by going backward from the Euler-Lagrange equation \ref{equ}
\begin{equation}
\label{lagrange}
\mathcal{L}=\frac{1}{2}\bigg [(\partial_\omega \varphi )^2-(\partial_{{{k^\mu}}} \varphi )^2-l^2 \varphi^2\bigg],
\end{equation}
The corresponding Hamiltonian is then given by
\begin{equation}
\label{hamiltonian}
\tilde{H}=\frac{1}{2}(\partial_{k^\mu} \varphi )^2+\frac{1}{2}\Pi^2+\frac{1}{2}l^2 \varphi^2,
\end{equation}
where $\Pi(\omega, k^\mu)$ is the conjugate momentum for $\varphi(\omega, k^\mu)$ defined as
\begin{equation}
\label{momentum}
\Pi({\omega, k^\mu})= \partial_{\omega} \varphi({\omega, k^\mu}),
\end{equation}
In what follows we will use the notation $\textbf{\emph{x}}$ and $\textbf{\emph{k}}$ for our space and momentum vector, respectively. The equation \ref{equ} has a general solution of the form
\begin{equation}
\label{solution0}
\varphi (\omega, \textbf{\emph{k}})=\int d^4{\emph{x}} \bigg [a_{\textbf{\emph{x}}} u_{\textbf{\emph{x}}} (\omega,\textbf{\emph{k}})+a^\dagger_{\textbf{\emph{x}}}u^*_{\textbf{\emph{x}}}(\omega,\textbf{\emph{k}})\bigg],
\end{equation}
By letting the canonical equal-energy commutation relations be
\begin{equation}
\label{commutation1}
[\varphi(\omega,{\textbf{\emph{k}}}), \varphi(\omega,{\textbf{\emph{k}}}')]= [\Pi(\omega,{\textbf{\emph{k}}}), \Pi(\omega,{\textbf{\emph{k}}}')]=0,
\end{equation}
\begin{equation}
\label{commutation2}
[\varphi(\omega,{\textbf{\emph{k}}}), \Pi(\omega,{\textbf{\emph{k}}}')]=i\delta({\textbf{\emph{k}}}-{\textbf{\emph{k}}}'),
\end{equation}
the creation and annihilation operators satisfy
\begin{equation}
\label{commutation}
 [a_{\textbf{\emph{x}}},a^\dagger _{\textbf{\emph{x}}'}]=\delta({\textbf{\emph{x}}}-{\textbf{\emph{x}}}'),  \enspace [a_{\textbf{\emph{x}}},a_{\textbf{\emph{x}}'}]= [a^\dagger_{\textbf{\emph{x}}},a^\dagger_{\textbf{\emph{x}}'}]=0,
\end{equation}
Thus, the scalar field operator $\varphi (\omega, \textbf{\emph{k}})$ in momentum space, can be represented in the form
\begin{equation}
\label{solution}
\varphi(\omega, \textbf{\emph{k}})=\frac{1}{(2\pi)^{2}\sqrt{x_0}}\int d^4{\emph{x}} \bigg [a_{\textbf{\emph{x}}}e^{i(\textbf{\emph{k}}.\textbf{\emph{x}}-\omega x_0)}+a^\dagger_{\textbf{\emph{x}}}e^{-i(\textbf{\emph{k}}.\textbf{\emph{x}}-\omega x_0)}\bigg],
\end{equation}
$u_{\textbf{\emph{x}}}=e^{i(\textbf{\emph{k}}.\textbf{\emph{x}}-\omega x_0)}$ and $u^*_{\textbf{\emph{x}}}=e^{-i(\textbf{\emph{k}}.\textbf{\emph{x}}-\omega x_0)}$ are identified as positive and negative ``frequency" solutions, respectively. That is, the ``frequency" is defined as $x_0=\sqrt{{\textbf{\emph{x}}}^2+l^2}$. $a_{\textbf{\emph{x}}}$ and $a^\dagger_{\textbf{\emph{x}}}$  are, respectively, the annihilation and creation operators that annihilates and creates excitations at spacetime point $(x_0,{\textbf{\emph{x}}})$. The associated vacuum states $|0\big>$ are defined by $a_{{\textbf{\emph{x}}}}|0\big>=0$ and $a^\dagger_{{\textbf{\emph{x}}}}|0\big>=|{\textbf{\emph{x}}}\big>$, respectively. The states $a^\dagger_{{\textbf{\emph{x}}}}|0\big>$ are interpreted as single particle states with time $x_0$ and position ${\textbf{\emph{x}}}$. One assumes, as seems reasonable since the wave equation is defined in momentum space, that the excitation ``carries" a specific $x_0$ and ${\textbf{\emph{x}}}$. Moreover, it is not difficult to notice that the excitations are located on the hyperboloid's upper sheet. This can be interpreted in the following way, ``spacetime excitations" carrying $x_0>0$ constitute the hyperboloid's upper sheet.\\\\
Establishing the theory in momentum space first, we shall now move on to coordinate space. To this end, we introduce a field $ \chi^{{AdS}}$ that we will call Anti-de Sitter field. The latter
 can be expanded in the basis $\{u_{\textbf{\emph{x}}}, u^*_{\textbf{\emph{x}}}\}$ as 
\begin{equation}
\label{KG}
 \chi^{{AdS}} (\omega, {\textbf{\emph{k}}})= \int d^4x \big [\psi_{\textbf{\emph{x}}} u_{\textbf{\emph{x}}}(\omega, {\textbf{\emph{k}}})+\psi^\dagger_{\textbf{\emph{x}}} u^*_{\textbf{\emph{x}}}(\omega, {\textbf{\emph{k}}})\big],
\end{equation}
where $\psi(\tau,{\textbf{\emph{x}}})=\int d^4k \big[a_{\textbf{\emph{k}}}\phi_{\textbf{\emph{k}}}(\tau, {\textbf{\emph{x}}})+\big(a_{\textbf{\emph{k}}}\big)^\dagger\phi^*_{\textbf{\emph{k}}}(\tau, {\textbf{\emph{x}}})\big]$ is a Klein-Gordon (K-G) field operator that satisfies the canonical commutation relations $[\psi(\tau,{\textbf{\emph{x}}}), \psi^\dagger(\tau,{\textbf{\emph{x'}}})]=\delta({\textbf{\emph{x}}}-{\textbf{\emph{x'}}})$ and $[\psi(\tau,{\textbf{\emph{x}}}), \Pi(\tau, {\textbf{\emph{x}}}')]=i\delta({\textbf{\emph{x}}}-{\textbf{\emph{x}}}')$, with ($\tau,{\textbf{\emph{x}}}$) being the usual Minkowski coordinates. On one hand, the operators $\psi_{{\textbf{\emph{x}}}}$ and $\psi^\dagger_{\textbf{\emph{x}}}$  act as Fourier coefficients in the field operator $\chi^{{AdS}}$'s expansion; and on the other hand, they act as Klein-Gordon annihilation and creation operators that annihilates/creates K-G particles at $(x_0,{\textbf{\emph{x}}})$. Note that the field $\zeta^{{AdS}}(x_0,{\textbf{\emph{x}}})=\psi_{{\textbf{\emph{x}}}} u_{{\textbf{\emph{x}}}}(\omega, {\textbf{\emph{k}}})+\psi^\dagger_{\textbf{\emph{x}}}u^*_{\textbf{\emph{x}}}(\omega,{\textbf{\emph{k}}})$ obeys the Klein-Gordon equation. \\\\
A more plausible way to write down the field $\zeta^{{AdS}}$ is to expand it in terms of the pair $\{\phi_{\textbf{\emph{k}}},\phi^*_{\textbf{\emph{k}}}\}_{_{\tau \to x_{0}}}$
\begin{equation}
\label{KG2}
 \zeta^{{AdS}}(x_0, {\textbf{\emph{x}}})=\int d^4k \big[a^{^{AdS}}_{\textbf{\emph{k}}}\phi_{\textbf{\emph{k}}}(x_0, {\textbf{\emph{x}}})+\big(a^{^{AdS}}_{\textbf{\emph{k}}}\big)^\dagger\phi^*_{\textbf{\emph{k}}}(x_0, {\textbf{\emph{x}}})\big],
\end{equation}
$a_{\textbf{\emph{k}}}^{^{AdS}}$ and $\big(a^{^{AdS}}_{\textbf{\emph{k}}}\big)^\dagger$ are annihilation and creation operators with respect to the modes $\phi_{\textbf{\emph{k}}}$ and $\phi^*_{\textbf{\emph{k}}}$. The corresponding vacuum states are then defined as $a_{\textbf{\emph{k}}}^{^{AdS}}|0\big>_{AdS}=0$ and $(a_{\textbf{\emph{k}}}^{^{AdS}})^\dagger=|{\textbf{\emph{k}}}\big>_{AdS}$, respectively. Note that the excitations associated to $(a_{\textbf{\emph{k}}}^{^{AdS}})^\dagger|0\big>_{AdS}$ carry an energy $\omega$ while those corresponding to $\zeta^{{AdS}}|0\big>_{AdS}$ are located at $x_0=\sqrt{{\textbf{\emph{x}}}^2+l^2}$. Thus, the field operator is associated with single particles with ``length" $l$ and mass $m$ in case where the field is massive, i.e., $\omega=\sqrt{{\textbf{\emph{k}}}^2+m^2}$. Given that the field operators in momentum space are related to those in coordinate space by Fourier transform, one may intuitively assume that excitations carrying a specific $\omega$ and ${\textbf{\emph{k}}}$ in coordinate space correspond to excitations that carry  a specific $x_0$ and {\textbf{\emph{x}}} in momentum space. Thus, we may use the terminologies particles and spacetime quanta interchangeably.\\\\
For more details on the quantization procedure, the reader is referred to Refs.\cite{35}$-$\cite{36}. Since we are interested in Gibbons-Hawking temperature, the thermal effect of anti-de Sitter hyperbolic geometry was computed by means of the field correlation function, following mostly Ref. \cite{36}. Thus, using Eq. \ref{KG2} the two-time correlation function ${{\big<\zeta^{{AdS}}(x_0)\zeta^{{AdS}}(x'_0)\big>}_{0_{AdS}}}$ for a massless field in the vacuum state $|0\big>_{AdS}$ reads\\
\begin{equation}
\label{correlation1}
{{\big<\zeta^{{AdS}}(x_0)\zeta^{{AdS}}(x'_0)\big>}_{0_{AdS}}}=\frac{1}{\pi}\frac{1} {\Delta {x_0}^2-\sum\limits_{\mu=1}^{4}\Delta x_\mu \Delta x^\mu},
\end{equation}
where $\Delta x_0=x'_0(t')-x_0(t)$ and $\Delta x_\mu=x'_\mu(t')-x_\mu(t)$. From the relations Eqs. $\ref{static coord1}-\ref{static coord3}$ it follows that
\begin{equation}
\label{difference}
\Delta {x_0}^2-\sum\limits_{\mu=1}^{4}\Delta  x_\mu \Delta x^\mu=-4l^2 sinh^2\bigg (\frac{t'-t}{2l}\bigg ),
\end{equation}
where we have made the approximation $r<<l$ so that the metric \ref{de sitter metric} reduces to the conventional Friedmann-Lema\^{i}tre-Robertson-Walker (FLRW) metric describing our universe in its static form. It turns out that the correlation function ${{\big<\zeta^{{AdS}}(x_0)\zeta^{{AdS}}(x'_0)\big>}_{0_{AdS}}}$ \ref{correlation1} takes the form
\begin{equation}
\label{correlation2}
{{\big<\zeta^{{AdS}}(x_0)\zeta^{{AdS}}(x'_0)\big>}_{0_{AdS}}}=-\frac{1}{4\pi l^2} csch^2\bigg (\frac{t'-t}{2l}\bigg ),
\end{equation}
That is a thermal-field correlation function with temperature $T=\frac{1}{2\pi l}$ which is nothing more than Gibbons-Hawking temperature.\\\\
It is worth mentioning that the relation \ref{correlation2} gives a rigorous physical meaning: The vacuum's thermal fluctuations arise due to the presence of anti-de Sitter field $\zeta^{AdS}$ whose quanta ``carry" a ``length" $l$. In the spirit of our analysis, Gibbons-Hawking radiation is an intrinsic property of the hyperbolic geometry and not of the tunnellling particle.  It might be known that the complex path method fails markedly to point out this feature, in the sense that the computations
depend strongly on the tunnelling particle (particularly on its spin). Nonetheless, it has been proved for many backgrounds that the tunnelling of particles
with different spins (s = 0, 1/2, 1 and s = 3/2) always yields the same temperature \cite{38}.\\\\
The results obtained till now, give rise to the following questions: Is the observed cosmic background radiation (CMB) nothing but the Gibbons-Hawking radiation? Is it the result of spacetime creation from nothing? Can inflation be generated by the creation of spacetime quanta? answering these questions is beyond the scope of this paper. They, the questions will be addressed later on and they will be explored further in an upcoming paper.\\\\

\section{q-INFLATIONARY MODEL}
This section is dedicated to study the properties (i.e. power spectrum and spectral index) of  large scale vacuum fluctuations in the q-inflationary scenario. The latter, which depends on the non extensive parameter $q$ measuring the deviation from the usual exponential expansion, is a strategy to (i) skirt scalar fields with fine-tuned potentials, (ii) skirt an exponential expansion that lasts forever and (iii) ensure a nearly but not perfectly scale-invariant spectrum, in agreement with observational data.\\\\
Dealing again with the metric
\begin{equation}
\label{anti de sitter metric}
ds^2=-\bigg (1+\frac{r^2}{l^2}\bigg)dt^2-\bigg[\frac{1}{1+\frac{r^2}{l^2}}dr^2+r^2(d\theta^2+sin^2\theta d\phi^2)\bigg],
\end{equation}
then making the following transformations \cite{39}
\begin{equation}
\label{anti de sitter coordinates transf}
\tilde{r}=\frac{a_i^{-1}r}{\sqrt{1+r^2/l^2}}e^{-t/l}, \qquad \tilde{t}=t+\frac{l}{2}ln\bigg(1+\frac{r^2}{l^2}\bigg), \qquad \tilde{\theta}=\theta, \qquad \tilde{\phi}=\phi,
\end{equation}
leads, under the Wick rotation mentioned in Section II, to the line element
\begin{equation}
\label{transf metric1}
ds^2=d\tilde{t}^2-a_i^{2}e^{2\tilde{t}/l}\big[d\tilde{r}^2+\tilde{r}^2d\tilde{\theta}^2+\tilde{r}^2sin^2\tilde{\theta}d\tilde{\phi}^2\big],
\end{equation}
 where $a_i$  is a constant denoting the initial value of  the scale factor $a(\tilde{t})=a_i e^{\tilde{t}/l}$ at $\tilde{t}=0$. Finally, introducing the space coordinates $\tilde{x}^1, \tilde{x}^2$ and $\tilde{x}^3$ which are related to $\tilde{r}, \tilde{\theta}$ and $\tilde{\phi}$ by the usual equations connecting Cartesian coordinates and polar coordinates in Euclidean space, Eq.\ref{transf metric1} may be written as
\begin{equation}
\label{transf metric2}
ds^2=d\tilde{t}^2-a_i^{2}e^{2\tilde{t}/l}d\tilde{\textbf{\emph{x}}}^2,
\end{equation}
Now let the scale factor $a$ being generalized to $a(\tilde{t})=a_i\bigg[1+(1-q)\sqrt{\frac{\Lambda}{3}}{\tilde{t}}\bigg]^{\frac{1}{1-q}}$ (provided that $q\neq1$), with $\sqrt{\frac{\Lambda}{3}}=\frac{1}{l}$. From now on, we will call the mentioned scale factor, a q-scale factor that denotes a q-expansion of the space concerned. It should be noted that in the limiting case $q\to1$, the usual exponential expansion is recovered. To restrict ourselves to an accelerating universe, values of $q>2$ are excluded since they correspond to a decelerating universe.\\\\ Based on these assumptions, the line element \ref{transf metric2} takes the following form
\begin{equation}
\label{tsallis transf metric2}
ds^2=d\tilde{t}^2-a_i^{2}\bigg[1+(1-q){\sqrt{\frac{\Lambda}{3}}{\tilde{t}}}\bigg]^{\frac{2}{1-q}}d\tilde{\textbf{\emph{x}}}^2,
\end{equation}
where the metric tensor
\begin{equation}
g_{\alpha \beta} = \begin{bmatrix}
       1 &0 & 0  & 0         \\[0.3em]
      0 &- a_i\bigg[1+(1-q){\sqrt{\frac{\Lambda}{3}}{\tilde{t}}}\bigg]^{\frac{1}{1-q}}&0&0\\[0.3em]
     0 &0& -a_i\bigg[1+(1-q){\sqrt{\frac{\Lambda}{3}}{\tilde{t}}}\bigg]^{\frac{1}{1-q}}&0\\[0.3em]
      0 &0&0& -a_i\bigg[1+(1-q){\sqrt{\frac{\Lambda}{3}}{\tilde{t}}}\bigg]^{\frac{1}{1-q}}
     \end{bmatrix}
\end{equation}
is diagonal. Note the only Riemann tensor's covariant derivatives that survive are $R_{i0i0;0}=R_{0i0i;0}=-R_{i00i;0}=-R_{0ii0;0}=\frac{1}{2}\frac{\partial^3g_{ii}}{\partial {\tilde{t}}^3}$ (with $i=1, 2, 3$), so that one can easily check that the Bianchi identities, i.e. $R_{\alpha \beta \gamma \delta; \lambda}+R_{\alpha \beta \lambda \gamma; \delta}+R_{\alpha \beta \delta \lambda; \gamma}=0$ still hold when incorporating the non-extensive parameter $q$.\\\\
From Einstein's field  equations
\begin{equation}
\label{einstein equations}
R_{\alpha \beta}-\frac{1}{2}g_{\alpha\beta} \mathcal{R}=8\pi GT_{\alpha\beta}
\end{equation}
with $T^\alpha_\beta={\rm{diag}}[\bar{\rho}, -\bar{p}, -\bar{p} ,-\bar{p}]$, one can get the corresponding deformed matter sector defined by an energy density
\begin{equation}
\label{deformed density}
\bar{\rho}=\frac{3}{8\pi G}{\frac{\Lambda}{3}}\bigg[1+(1-q)\sqrt{\frac{\Lambda}{3}} \tilde{t}\bigg]^{-2}
\end{equation}
and a pressure
\begin{equation}
\label{deformed pressure}
\bar{p}=-(2q+1)\frac{1}{8\pi G}{\frac{\Lambda}{3}}\bigg[1+(1-q)\sqrt{\frac{\Lambda}{3}} \tilde{t}\bigg]^{-2}=-(2q+1)\frac{\bar{\rho}}{3}
\end{equation}
where during inflation ordinary matter sector is negligible compared to the component triggering inflation. It is easily checked that, in the limiting case $q\to 1$, the standard relations $\bar{\rho}=\frac{\Lambda}{8\pi G}$ and $\bar{p}=-\bar{\rho}$, for a cosmological constant $\Lambda$, are recovered. It is worth noting that the relations \ref{deformed density} and \ref{deformed pressure} fulfill the conservation equations for the energy-momentum tensor $T^{\alpha \beta}$ given by $T^{\alpha \beta}_{;\beta}=0$. That is
\begin{equation}
\label{conservation eq}
\dot{\bar{\rho}}+3H(\bar{\rho}+\bar{p})=0,
\end{equation}
where $H$ is the expansion rate defined as $H=\frac{\dot{a}}{a}$ (with over-dot being the derivative over time $\tilde{t}$).\\
We consider now the dynamics of a scalar fluctuations $\delta \zeta^{^{{AdS}}}(\tilde{\eta},\tilde{\textbf{\emph{x}}})$ of our previously introduced field $\zeta^{^{{AdS}}}$, in the new coordinate system $(\tilde{t},\tilde{\textbf{\emph{x}}})$. Expanding the scalar fluctuations $\delta \zeta^{^{{AdS}}}$, which will be our inflaton field, in Fourier modes yields 
\begin{equation}
\label{massless scalar filed}
\delta\zeta^{^{{AdS}}}(\tilde{\eta},\tilde{\textbf{\emph{x}}}) =\int d\textbf{\emph{k}} [a^{AdS}_{\textbf{\emph{k}}}e^{i\textbf{\emph{k}}.\tilde{\textbf{\emph{x}}}}\delta v_{\textbf{\emph{k}}}(\tilde{\eta})+(a^{AdS}_{\textbf{\emph{k}}})^\dagger e^{-i\textbf{\emph{k}}.\tilde{\textbf{\emph{x}}}}\delta v^*_{\textbf{\emph{k}}}(\tilde{\eta})]
\end{equation}
where $\tilde{\eta}$ is the conformal time, from here on we omit the superscript $``AdS"$ in order to maintain simple notations. In an inflationary background, the mode function $\delta v_\textbf{\emph{k}}$ satisfies the following equation
\begin{equation}
\label{expanding kg}
\delta v_{\textbf{\emph{k}} }^{''}+2\mathcal{H}\delta v_{\textbf{\emph{k}} }^{'}+k^2\delta v_{\textbf{\emph{k}}}+a^2\frac{\partial{V}}{\partial{\delta v_{\textbf{\emph{k}}}}}=0,
\end{equation}
where the prime denotes derivative with respect to the conformal time $\tilde{\eta}$, $\mathcal{H}=a'/a$ and $V$ is the potential of the scalar field. Firstly, we consider the simple case of an exponential expansion ($q\to1$), for which $H=\sqrt{\frac{\Lambda}{3}}$ is time-independent and $a(\tilde{\eta})=-\frac{1}{H\tilde{\eta}}$. \\
Equation. \ref{expanding kg} can be recast into
\begin{equation}
\label{recasted expanding kg1}
\delta \sigma_{\textbf{\emph{k}} }^{''}+\bigg(k^2-\frac{a''}{a} +m_\zeta^2a^2 \bigg)\delta \sigma_{\textbf{\emph{k}} }=0,
\end{equation}
\begin{equation}
\label{recasted expanding kg2}
\delta \sigma_{\textbf{\emph{k}} }^{''}+\bigg[k^2-\frac{1}{\tilde{\eta}^2}\big(\nu^2-\frac{1}{4}-\frac{m^2_\zeta}{H^2}\big)\bigg]\delta \sigma_{\textbf{\emph{k}} }=0,
\end{equation}
with $\sigma=a(\tilde{\eta})v$, $m_\zeta$ being the mass of the scalar field and and $\nu^2=\frac{9}{4}$. In the following, we will neglect the last term of Eq. \ref{recasted expanding kg2} by taking into account the assumption $\frac{m_\zeta^2}{2H^2}<<1$. \\ The generic solution to Eq. \ref{recasted expanding kg1} is 
\begin{equation}
\label{generic solution}
\delta \sigma_{\textbf{\emph{k}} }=\sqrt{-\tilde{\eta}}\big[c_1(k)H^{(1)}_\nu(-k\tilde{\eta})+c_2H^{(2)}_\nu(-k\tilde{\eta})\big],
\end{equation}
where $H^{(1)}_\nu$ and $H^{(2)}_\nu$ are the Hankel's functions of the first and second kind, respectively. Imposing that in the ultraviolet regime $k>>aH$, equation.\ref{recasted expanding kg1} admits a plane wave solution $e^{-ik\tilde{\eta}}/\sqrt{2k}$ that is expected in flat spacetime and having the following known form in hand
\begin{equation}
\label{henkel}
H^{(1)}_\nu(-k\tilde{\eta}>>1) \sim \sqrt{-\frac{2}{\pi k\tilde{\eta}}}e^{i(-k\tilde{\eta}-\frac{\pi}{2}\nu-\frac{\pi}{4})},
\end{equation}
we set $c_2(k)=0$ and $c_1(k)=\frac{\sqrt{\pi}}{2}e^{i(\nu+\frac{1}{2})\frac{\pi}{2}}$. The exact solution then reads
\begin{equation}
\label{exact henkel}
\delta \sigma_{\textbf{\emph{k}} }=\frac{\sqrt{\pi}}{2}e^{i(\nu+\frac{1}{2})\frac{\pi}{2}}\sqrt{-\tilde{\eta}}H^{(1)}_\nu(-k\tilde{\eta}),
\end{equation}
The expansion quickly redshifts short-wavelength vacuum fluctuations until their wavelengths go beyond the horizon size $H^{-1}$, hence the quantum modes cease to evolve and ``freeze out" as classical fluctuations.  Here we are interested in fluctuations of wavelengths that are much larger than the Hubble horizon (i.e. $\emph{k}<<aH$). Thus, on super-horizon scales we have $H^{(1)}_\nu(-k\tilde{\eta}<<1)\sim \sqrt{2/\pi}e^{-i\frac{\pi}{2}}2^{\nu-\frac{3}{2}}(\Gamma(\nu)/\Gamma(3/2))(-k\tilde{\eta})^{-\nu}$, which in turn yields
\begin{equation}
\label{superhorizon exact henkel}
\delta \sigma_{\textbf{\emph{k}} }=e^{i(\nu-\frac{1}{2})\frac{\pi}{2}}2^{\nu-\frac{3}{2}}\frac{\Gamma(\nu)}{\Gamma(3/2)}\frac{1}{\sqrt{2k}}(-k\tilde{\eta})^{\frac{1}{2}-\nu},
\end{equation}
Going back to the variable $\delta v_{\textbf{\emph{k}} }$, the fluctuation on super-horizon scales is given by
\begin{equation}
\label{solution of expanding kg}
\delta v_{\textbf{\emph{k}} }\simeq\frac{H} {\sqrt{2{\emph{k}}^3}}\bigg(\frac{\emph{k}}{aH}\bigg)^{{\frac{3}{2}-\nu}},
\end{equation}
which, for $\nu=\frac{3}{2}$, reads
\begin{equation}
\label{super horizon solution}
\delta v_{\textbf{\emph{k}} }\simeq\frac{H}{\sqrt{2\emph{k}^3}},
\end{equation}
A useful quantity to describe the perturbations properties is the so-called  power spectrum. It is derived from the average amplitude of the inflationary perturbations
\begin{equation}
\label{power spectrum1}
\big<0|{\delta\zeta}^\dagger(\tilde{t},\tilde{\textbf{\emph{x}}}) \delta\zeta(\tilde{t},\tilde{\textbf{\emph{x}}})|0\big>=\int \frac{d\emph{k}}{k}\frac{\emph{k}^3}{2\pi^2}|\delta v_\textbf{\emph{k}} |^2,
\end{equation}
Thus, the power spectrum of vacuum fluctuations is defined as
\begin{equation}
\label{power spectrum2}
 \mathcal{P}_{\emph{k}}= \frac{\emph{k}^3}{2\pi^2}|\delta v_\textbf{\emph{k}} |^2=\bigg(\frac{H}{{2\pi}}\bigg)^2,
\end{equation}
The spectral index $n_s$ is calculated throughout the logarithmic derivative of the power spectrum
\begin{equation}
\label{spectral index1}
n_s-1=\frac {dln \mathcal{P}_{\emph{k}}}{dln{\emph{k}}},
\end{equation}
which yields $n_s=1$ for an exponential expansion where $H=\sqrt{\frac{\Lambda}{3}}$. This corresponds to a scale-invariant spectrum.\\
For $q\neq 1$ however, the expansion rate $H$ is given by
\begin{equation}
\label{tasllis H}
H=\sqrt{\frac{\Lambda}{3}}\bigg[1+(1-q)\sqrt{\frac{\Lambda}{3}} \tilde{t}\bigg]^{-1},
\end{equation}
while the term $a''/a$ appearing in Eq. \ref{recasted expanding kg1} reads
\begin{equation}
\label{scale a}
\frac{a''}{a}=\frac{2-\epsilon}{\tilde{\eta}^2(1-\epsilon)^2}\simeq\frac{1}{\tilde{\eta}^2}(2+3\epsilon), 
\end{equation}
where we have used the fact that, for a Hubble rate that changes with time as $\dot{H}=-(1-q)H^2$, the scale factor, for small values of $\epsilon=1-q$, takes the form $a=-\frac{1}{H\tilde{\eta}}\frac{1}{1-\epsilon}$. Substituting Eq. \ref{scale a} into Eq. \ref{recasted expanding kg1}, we get Eq. \ref{recasted expanding kg2} with $\nu \simeq \frac{3}{2}+\epsilon$. So that, from Eq. \ref{solution of expanding kg}, the solution takes the form
\begin{equation}
\label{quasi de sitter solution}
\delta v_{\emph{k}}\simeq\frac{H}{\sqrt{2{\emph{k}}^3}}\bigg(\frac{\emph{k}}{aH}\bigg)^{-(1-q)},
\end{equation}
It turns out that the corresponding power spectrum, making use of Eq. \ref{power spectrum1}, reads
\begin{equation}
\label{tasllis power spectrum}
\mathcal{P}_{\emph{k}}=\bigg(\frac{H}{{2\pi}}\bigg)^2\bigg(\frac{\emph{k}}{aH}\bigg)^{-2(1-q)}=\bigg(\frac{1}{2\pi}\sqrt{\frac{\Lambda}{3}}\bigg[1+(1-q)\sqrt{\frac{\Lambda}{3}} \tilde{t}\bigg]^{-1}\bigg)^2\bigg(\frac{\emph{k}}{{aH}}\bigg)^{-2(1-q)},
\end{equation}
It is easily checked, from Eq. \ref{tasllis power spectrum}, that the spectrum of the Bunch$-$Davies vacuum \ref{power spectrum2} is recovered in the limiting case $q\to1$. On the other hand, the spectrum's amplitude is slightly diminished regarding that of the Bunch$-$Davies vacuum for $\tilde{t}\sim \sqrt{\frac{3}{\Lambda}}$. However, the amplitude is drastically diminished for $\tilde{t}>>\sqrt{\frac{3}{\Lambda}}$, which means that the effect of the parameter $q$ is appreciable only at late times, i.e., end of inflation.\\\\
As a side note, inflation is due to the ``cosmological constant" $\sqrt{{{\Lambda}}/{3}}={1}/{l}$, which  is nothing more than the inverse of the hyperboloid curvature's radius. The physical meaning of this result, from the standpoint of spacetime quantization (i.e. $x_0^2=\textbf{\emph{x}}^2+l^2$), is that spacetime excitations (see Sec. II) ``carrying" a ``length" $l$ or a ``curvature" $1/l^2$  are implicitly triggering the q-exponential inflation. Furthermore, there would be no inflation regardless of  its nature (exponential or linear..) if there was no excitation, i.e., $x_0=0$.  \\
\section{OBSERVATIONAL CONSTRAINTS ON THE q-PARAMETER}
The value of the spectral index measured  by the Planck collaboration 
including the results of WMAP and those based on
the investigation of baryon acoustic oscillations (BAO)
$\cite{40}-\cite{42}$, is required to lie in the range $[0.955-0.98]$ at $95\%$ CL, which rules out the scale invariant spectrum (i.e. $n_s=1$) at more than 5$\sigma$ confidence level. Based on this finding, we can safely discard inflationary models for which $q=1$. Hence, we must pick out the range within which the value of $q$ must lie.
For such a purpose, let us put $\zeta(\tilde{t},\tilde{\textbf{\emph{x}}})=\zeta_0(\tilde{t})+\delta \zeta(\tilde{t},\tilde{\textbf{\emph{x}}})$, where $\zeta_0(\tilde{t})$ represents the homogeneous background part of the field whereas $\delta \zeta(\tilde{t},\tilde{\textbf{\emph{x}}})$ denotes the spatially fluctuating part. During inflation, the universe was assumed to be in a Friedmann-Lema\^{i}tre-Robertson-Walker state with small inhomogeneities. Roughly speaking, scalar perturbations about a homogeneous universe filled with an energy density can be described by perturbations in the energy density as well as perturbations in the metric
\begin{equation}
\label{metric perturabtion}
ds^2=(1+2\Phi)d\tilde{t}^2-a^2(\tilde{t}) (1-2\Phi)[d\tilde{r}^2+\tilde{r}^2(d\tilde{\theta}^2+sin^2\tilde{\theta} d\tilde{\phi}^2)],
\end{equation}
where we have considered only scalar perturbations $\Phi(\tilde{t},\tilde{\textbf{\emph{x}}})$. The latter plays a similar role as that of the Newtonian potential used to describe weak gravitational fields (A. Linde, 2005, p.176) \footnote{See A. D. Linde, Particle physics and inflationary cosmology, Contemp. Concepts Phys.5(2005)1-362.} (i.e. Schwarzschild metric). $a(\tilde{t})$ is the q-scale factor described above.\\
A useful gauge-invariant quantity for characterizing scalar perturbations during inflation is the curvature perturbation $\mathcal{R}$ defined as 
\begin{equation}
\label{comoving curvature perturbation}
\mathcal{R}=\Phi+H\frac{\delta \zeta}{\dot{\zeta}_0},
\end{equation}
Working in the spatially flat gauge (i.e. $\Phi=0$), the corresponding dimensionless power spectrum reads
\begin{equation}
\label{comoving curvature perturbation1}
\mathcal{P_R}\bigg|_{\Phi=0}=\frac{H^2}{\dot{\zeta}_0^{^{^2}}}\bigg({\frac{H}{2\pi}}\bigg)^2\bigg(\frac{\emph{k}}{aH}\bigg)^{-2(1-q)},
\end{equation}
Since the scalar field $\zeta_0$ is our inflaton field, it must obey the equation of state given in Eq. \ref{deformed pressure}. Consequently, the energy density and pressure due to  $\zeta_0$ are given by the following relations
\begin{equation}
\label{field density}
\bar{\rho} = {\dot{\zeta}_0}^{^{^2}}/2+V({\zeta}_0),
\end{equation}
\begin{equation}
\label{field pressure}
\bar{p}= {\dot{\zeta}_0}^{^{^2}}/2-V({\zeta}_0),
\end{equation}
Thus, one can get, using the above relations along with Eq. \ref{deformed pressure} the form of  ${\dot{\zeta}_0}$
\begin{equation}
\label{field dot}
{\dot{\zeta}_0}^{^{^2}}=\frac{2(1-q)}{3}\bar{\rho}
\end{equation}
where $\bar{\rho}=\frac{3 m_p^2}{8\pi} H^2$ is the homogeneous background energy density, it should be noted that the fluctuations in $T_{\mu\nu}$, have been considered to be negligible compared with the energy density $\bar{\rho}$.  
Substituting the relation \ref{field dot} into Eq.\ref{comoving curvature perturbation1}, the power spectrum $\mathcal{P_R}$ becomes
\begin{equation}
\label{comoving curvature perturbation2}
\mathcal{P_R}\bigg|_{\Phi=0}=\frac{4\pi}{(1-q)m_p^2}\bigg({\frac{H}{2\pi}}\bigg)^2\bigg(\frac{\emph{k}}{aH}\bigg)^{-2(1-q)}=\mathcal{A_R}\bigg(\frac{\emph{k}}{aH}\bigg)^{-2(1-q)},
\end{equation}
where $m_p=\frac{1}{\sqrt{G}}$ is the Planck mass and $\mathcal{A_R}$ being the amplitude of the power spectrum. The logarithmic derivative of $\mathcal{P_R}$ is then expressed as
\begin{equation}
\label{tasllis n}
n_{\mathcal{R}}-1=\frac {dln \mathcal{P_R}(\emph{k})}{dln{\emph{k}}}=-2(1-q),
\end{equation}
Since the observational range of $n_{\mathcal{R}}$ is $[0.955-0.98]$, $q$ is restricted to be within the range $0.9775\leq q \leq 0.99$. Furthermore, the tensor-to-scalar ratio $r$ is defined as
\begin{equation}
\label{tensor to scalar ratio}
r=\frac{\mathcal{P_T}}{\mathcal{P_R}}=\frac{16}{3}(1-q),
\end{equation}
where $\mathcal{P_T}=8.\frac{8\pi }{3m_p^2}\bigg({\frac{H}{2\pi}}\bigg)^2\bigg(\frac{\emph{k}}{aH}\bigg)^{-2(1-q)}$ is the well-known tensor perturbations spectrum. The above relation restricts the range of the tensor-to-scalar ratio to $0.053 \leq r \leq 0.12$, which is consistent with the limit $r<0.12$ (at 95\% CL) set by Planck including BAO data \cite{43}.\\\\
Another interesting feature of the inflationary cosmology is reheating. The latter describes production of the standard particles after the accelerated inflationary era where the universe has gone through supercooling. In the standard reheating process, the inflaton field decays perturbatively into a set of particles and it starts oscillating around the minimum of its potential during the decay mechanism \cite{44}. This reheating mechanism does not work for our proposed field. Nonetheless, reheating can be achieved if one assumes that, at the end of inflation, the energy density $\bar{\rho}$ is converted instantaneously into radiation $\bar{\rho}\sim \rho_r \sim \frac{\pi^2}{30}N(T)T^4$. It follows that the reheating temperature, the value of which can be computed at the end of inflation, takes the form
\begin{equation}
\label{reheating temp1}
T_{RH}\bigg|_{\tilde{t}=\tilde{t}_{end}}=\bigg[\frac{90}{8\pi^3 N(T_{RH})}m_p^2H^2_{end}\bigg]^{\frac{1}{4}},
\end{equation}
where $N(T_{RH})$ is the effective number of degrees of freedom at $T=T_{RH}$, with $N(T_{RH})\sim 10^2-10^3$.\\
Inspecting Eq. \ref{reheating temp1}, it is worth noting that the reheating temperature $T_{RH}$ and Gibbons-Hawking temperature $T_{GH}=\frac{1}{\pi}\sqrt{\frac{\Lambda}{12}}$ are related via
\begin{equation}
\label{reheating and Gibbons}
T_{RH}=\bigg[\frac{180\pi^2m_p^4l^4(1-q)e^{-2N_*(1-q)}}{N(T_{RH})}\mathcal{A}_{\mathcal{R}}\bigg]^{\frac{1}{4}}T_{GH},
\end{equation}
where we have used the fact that the Hubble scale at the end of inflation, $H_{end}$ , and the Hubble scale $H_*$, i.e. $H_*^2=\mathcal{A}_{\mathcal{R}}(1-q)m_p^2\pi$, at the pivot scale $k_*$ which leaves the horizon $N_*$ e-folds before the end of inflation, are related via $N_*=\int_{t_*}^{t_{end}} Hdt$, namely
\begin{equation}
\label{efold}
e^{N_*(1-q)}=\frac{H_*}{H_{end}},
\end{equation}
It should be noted that one can compute $N_*$,  assuming an instantaneous reheating, i.e. $\rho_{_{RH}}=\rho_{end}$, from the matching equation
\begin{equation}
\label{matching equation}
N_*=62.396+{\rm{ln}}\bigg(\frac{H_*}{m_p}\bigg)-\frac{1}{4}{\rm{ln}}\frac{\rho_{end}}{m^4_p}
\end{equation}
which is drawn from Eq. (20) of Ref. $\cite{45}$, with $\rho_{end}=\frac{3 m_p^2}{8\pi} H^2_{end}$. Combining Eqs. \ref{efold} and \ref{matching equation}, one can compute $N_*$ and hence derive $T_{RH}$ via Eq. \ref{reheating temp1}. Note that CMB data constraint $H_*^2/(1-q)$ through the amplitude of the anisotropies $\mathcal{A}_{\mathcal{R}}$, as well as $q$ from the spectral index. Consequently, one may expect CMB data to also provide some information on $N_*$ and $T_{RH}$.\\\\
In what follows we will examine the observational constraints
on the free parameters $\{q, l, t_*\}$ of the model, and the corresponding derived parameters $\{r, N_*, T_{RH}\}$.\\
\section{DATA ANALYSIS}
\begin{figure}
\begin{center}
\includegraphics[scale=0.3]{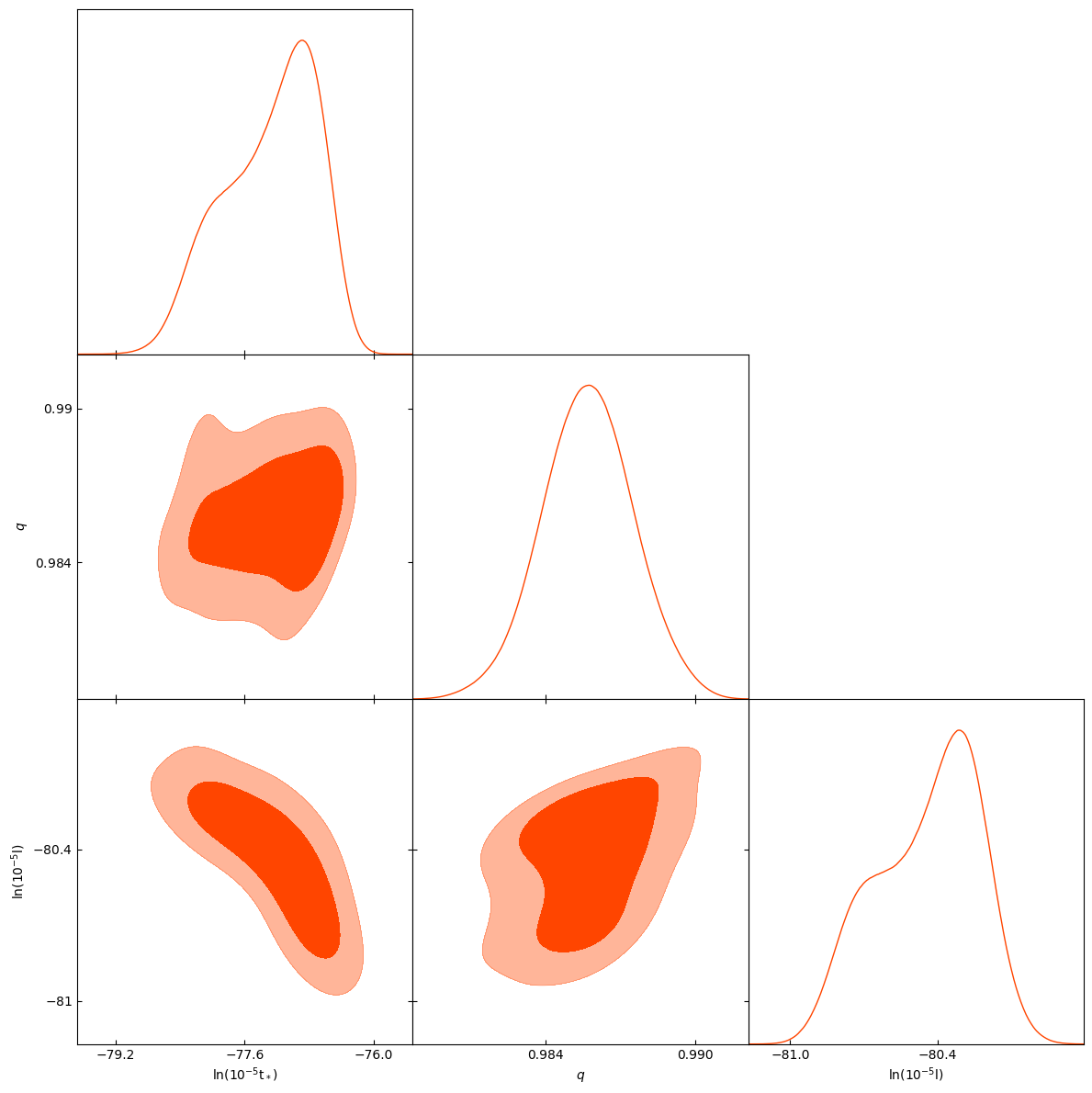}
\caption{2D joint posterior probability distributions and 1D  marginal posterior probability
distribution of the q-inflation parameters $q$, ${\rm{ln}}(10^{-5}l)$ and ${\rm{ln}}(10^{-5}t_*)$.}
\label{figure:bimodal}
\end{center}
\end{figure}

\begin{figure}
\begin{center}
\includegraphics[scale=0.3]{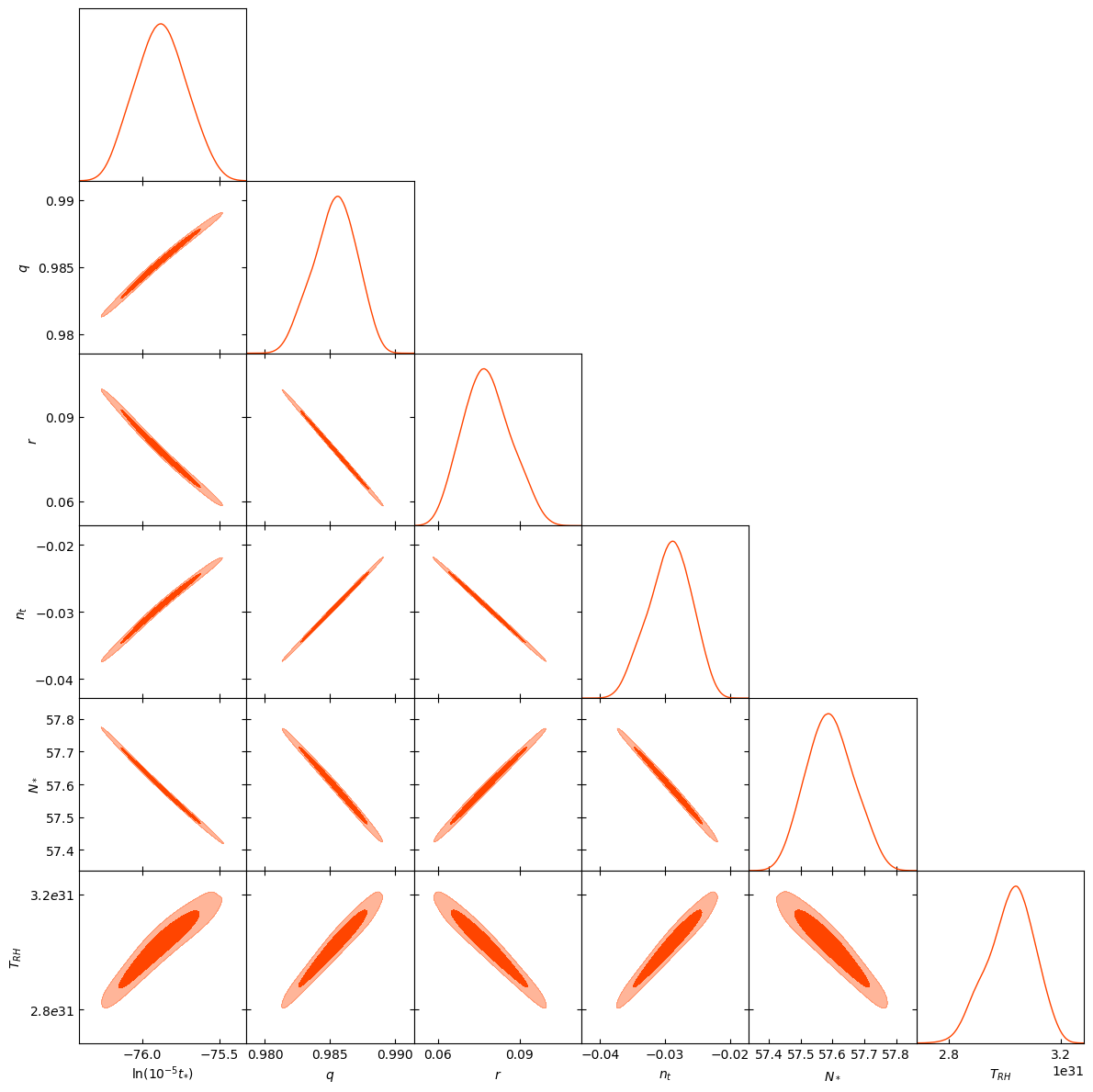}
\caption{2D joint posterior probability distributions and 1D  marginal posterior probability
distribution of the q-inflation parameters $q$ and ${\rm{ln}}(10^{-5}t_*)$ as well as related derived parameters ($r, n_t, N_*$ and $T_{RH}$).}
\label{figure:parameters}
\end{center}
\end{figure}

\begin{figure}
\begin{center}
\includegraphics[scale=0.35]{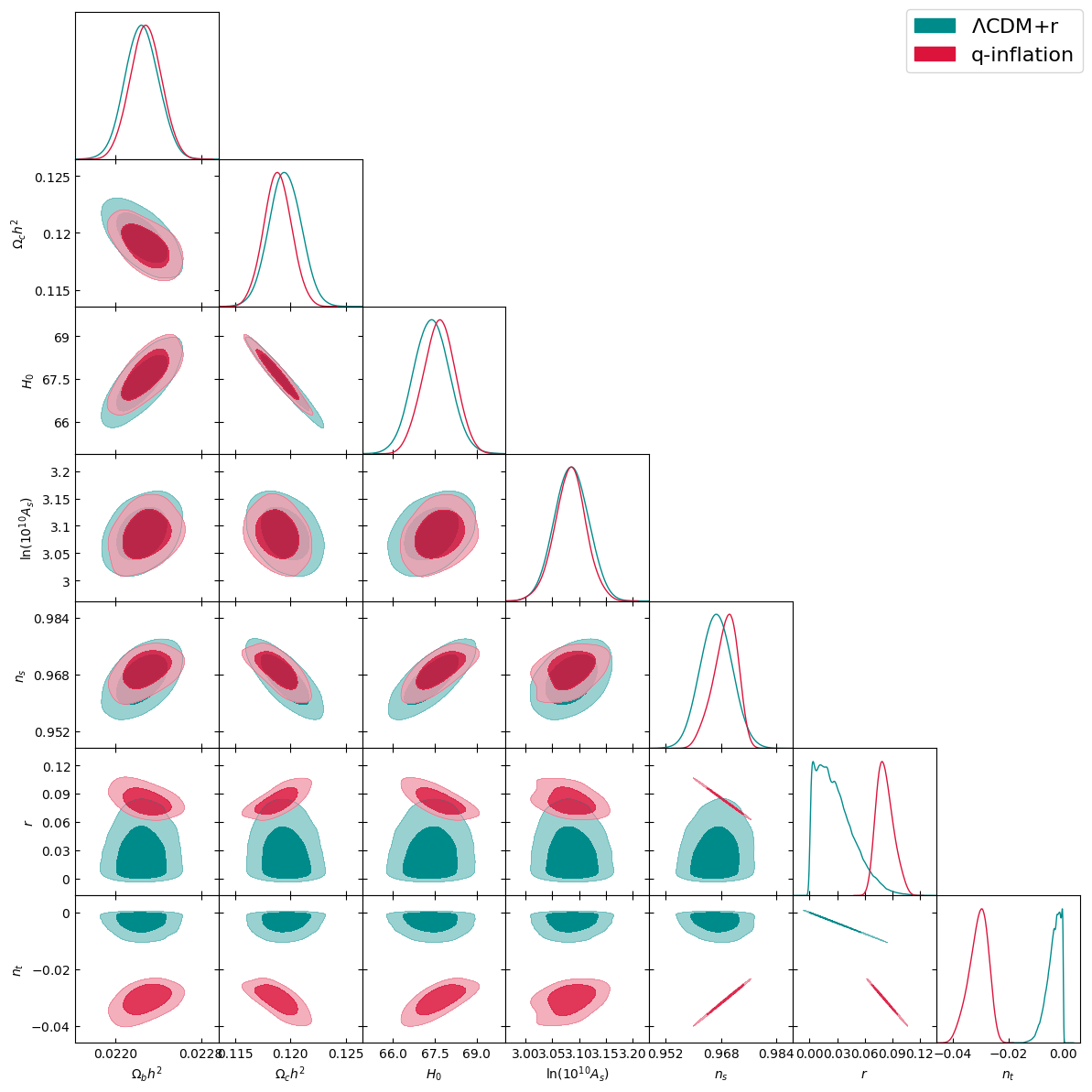}
\caption{Probability distribution functions and marginalized confidence regions at $68\%$ and $95\%$ for the parameters $\Omega_b h^2 , \Omega_c h^2, H_0, {\rm{ln}}(10^{10}A_s), n_s, r$ and $n_t$ of the q-inflation (crimson contours) and $\Lambda CDM+r$ model (cyan contours).}
\label{figure:comparaison parameters}
\end{center}
\end{figure}

In order to impose constraints on the q-inflationary model's parameters, we have used a modified version of the Boltzmann CAMB code $\cite{46}-\cite{48}$ and the Monte Carlo Markov Chain (MCMC) analysis $\cite{49}$ provided by the publicly available CosmoMC package\footnote{http://cosmologist.info/cosmomc/}. Therefore, the inflationary sector has been modified by plugging in the following power spectrum parameterizations 
\begin{equation}
\label{scalar camb}
\mathcal{P}_{\mathcal{R}}(k)=\frac{1}{(1-q)\pi m_p^2}{l^{-2}[1+(1-q)\tilde{t}_*/l]^{-2}}\bigg(\frac{k}{k_*}\bigg)^{-2(1-q)}
\end{equation}
\begin{equation}
\label{tensor camb}
\mathcal{P}_{t}(k)=r.\frac{1}{(1-q)\pi m_p^2}{l^{-2}[1+(1-q)\tilde{t}_*/l]^{-2}}\bigg(\frac{k}{k_*}\bigg)^{n_t}
\end{equation}
where $k_*$ denotes an arbitrary pivot scale and $n_t=-2(1-q)$ is the tilt of the tensor power spectrum. Thus, to constrain the model parameters \{$q$, $l$, $t_*$\}, and derive their respective posterior probability distributions, we have used Planck 2018 data set (TT, TE, EE+lowTEB) \cite{50} with low $l$ likelihood ($0\leq l \leq 29$) and high $l$ likelihood ($30\leq l \leq 2508$), estimated using commander. It is worth noting that, instead of $q, l$ and $\tilde{t}_*$, we have constrained $n_s=2q-1, {\rm{ln}}[10^{-5}l] $ and ${\rm{ln}}[10^{-5}\tilde{t}_*]$ in such away to use the standard parameterization provided in the CAMB code, i.e. $[l+(1-q)\tilde{t}_* ]\propto 1/\sqrt{A_s}$. The free parameters were set as uniform priors, along with the other parameters of the standard cosmological model:  baryon density ($\Omega_bh^2$), cold dark matter density ($\Omega_ch^2$), Thomson scattering optical depth due to re-ionization ($\tau$), and angular size of horizon ($\theta$). The priors are summarized in Table.\ref{table:parameters input}. Due to their negligible effect on the CMB power spectrum,  the effective number of neutrinos $N_\nu$, Helium mass fraction $Y_p$ and the width of re-ionization were kept fixed at their default values $3.046, 0.24$ and $0.5$ respectively. We have also fixed the pivot scale to $k_* = 0.05 \rm{Mpc}^{-1}$.\\\\
\begin{table}[h!]
\begin{center}
\label{table:parameters input}
\begin{tabular} {| l| c| c|}
\hline
 Parameter &  Lower limit & Upper limit\\
\hline 
{\emph{$\Omega_b h^2 $}} & $0.005$ & $0.1$\\
{\emph{$\Omega_c h^2 $}} & $0.001$ & $0.99$\\
{\emph{$\theta $}} & $0.5$ & $10$\\
{\emph{$\tau$}} & $0.01$ & $0.8$\\
{\emph{$n_s=2q-1$}} & $0.8$ & $1.2$\\[1ex]
{$\rm{ln}[10^{-5}\emph{$\tilde{t}_*$}]$}& $-90$ & $-74$\\
{$\rm{ln}[10^{-5}\emph{$l$}]$}& $-91.62$ & $-70$\\
\hline
\end{tabular}
\caption{Uniform priors used in MCMC parameters' estimation.}
\label{table:parameters input}
\end{center}
\end{table} 
It is noteworthy to mention that during the MCMC analysis we encountered a degeneracy problem in the parameter space $\{l, \tilde{t}_*\}$. This is illustrated in Fig. \ref{figure:bimodal} where the posterior probability distribution of the parameters concerned can exhibit multiple peaks and/or subpeaks describing a multi-modal pattern. Due to this degeneracy, CosmoMC sampling of the q-inflation undergoes a longer time to find the right mixing between parameters and thereby reaching the desired convergence is considerably a slack task. To overcome the problem of degeneracy between the parameters $l$ and $t_*$ it can be useful to simply discard the parameter $l$. Doing so is justified since for an extremely huge number of total e-folds the Hubble rate near to the end of inflation can be approximated by $H_*\simeq [(1-q)\tilde{t}_*]^{-1}$\footnote{At the end of inflation we can safely set $(1-q)\tilde{t}_{end}/l>>1$, so that $H_{end}\simeq [(1-q)\tilde{t}_{end}]^{-1}$}, with $\tilde{t}_*=e^{-(1-q)N_*}\tilde{t}_{end}$. In the following we will only discuss the results obtained from the latter case.\\\\
One-dimensional and two-dimensional marginalised posterior distributions for the model's parameters, are shown in Fig. \ref{figure:parameters}, while values at 95\% CL and best fit values are quoted in Table. \ref{table:data}. For comparative purposes, Table \ref{table:data} also displays the values at 95\% and the best fit for the standard normal inflation ($\Lambda CDM+r$). Furthermore, Figure \ref{figure:comparaison parameters} depicts the probability distribution functions and the marginalized confidence regions at $68\%$ and $95\%$ for the two models, where the crimson contours denote the constraints on the q-inflation while the cyan contours denote the constraints on $\Lambda CDM+r$ model. It is worth noting that, since they don't possess the same number of free parameters the two models are compared through the well-known Akaike information criterion (AIC) \cite{51}
\begin{equation}
{\rm{AIC}}=-2{\rm{ln}}\mathcal{L}_{max}+2N=\chi_{min}^2+2N,
\end{equation}
where $\mathcal{L}_{max}$ is the maximum likelihood and $N$ is the number of free parameters of the model. The difference in AIC values ($\Delta{\rm{AIC}} <2$) has an evidence support for the model under consideration.\\\\
Note that, unlike that of the standard normal inflation, the tensor-to-scalar ratio $r$ of the q-inflation is not considered as a free parameter but calculated as of function of $n_s$ throughout the relation \ref{tensor to scalar ratio}. Furthermore, a consistency relation between the tensor-to-scalar ratio $ r$ and the tilt of the tensor power spectrum $n_t$, namely $n_t=-3r/8$, has been used. The latter differs from the usual consistency relation by a factor of $3$. Since they are functions of a single free parameter $n_s$, $r$ and $n_t$ are strongly correlated and hence are more accurately constrained. The number of e-folds $N_*=57.60$  lies within the  range $50 < N_* < 60$ \cite{52} expected for an instantaneous reheating stage. Moreover, the reheating temperature $T_{RH}$ is found to be about $3.9\times10^{16} GeV$, and therefore inflation is halted at $\tilde{t}_{end}=8.5\times10^{-37} s$, from which it follows that $l<<3.78\times10^{-30}m$.\\\\
Apart from the rest of the models parameters, we see that the best fit tensor-to-scalar ratio $r$ is 0.079 for the q-inflation and 0.022 for the standard normal inflation. Thus, the magnitude of $r$ can be of help in distinguishing the standard normal inflation from the q-inflation. Meanwhile, the value of the tensor-to-scalar ratio cannot yet be determined precisely from the present Planck data. \\\\
\begin{savenotes}
\begin{table}[h!]

\begin{center}

\begin{tabular} { |l  |c| c|c|c|}
\hline
& \multicolumn{2}{|c}{q-inflation} & \multicolumn{2} {|c|}{$\Lambda CDM+r$} \\[1ex]
\hline 
 Parameter &  95\% limits &Best fit& 95\% limits & Best fit \\[1ex]
\hline
{$n_s            $} & $0.9707^{+0.0061}_{-0.0065}$ & $0.9705$ &$0.9666^{+0.0092}_{-0.0090}$ &$0.9670$\\[1ex]

{${\rm{ln}}[10^{-5}\tilde{t}_*] $} & $-75.88^{+0.32}_{-0.31}    $& $-75.90$ & $-$ & $-$\\[1ex]

$q                         $ &$0.9854^{+0.0031}_{-0.0033}$& $0.9852$ & $-$ & $-$\\[1ex]

$r                         $ & $0.078^{+0.017}_{-0.016}   $& $0.079$ & $0.029^{+0.040}_{-0.029}   $  &$0.022$\\[1ex]
$n_t                       $ & $-0.0293^{+0.0061}_{-0.0065}$ &$-0.0296$&$-0.0036^{+0.0037}_{-0.0050}$ &$-0.0028$\\[1ex]

${\rm{ln}}[10^{10}A_s]        $ & $3.087^{+0.062}_{-0.056}   $ & $3.083$ &$3.086^{+0.062}_{-0.062}   $  & $3.088$\\[1ex]

{$\Omega_b h^2   $} &  $0.02231^{+0.00030}_{-0.00029}$& $0.02225$ & $0.02224^{+0.00031}_{-0.00030}$ & $0.02226$\\[1ex]
{$\Omega_c h^2   $} & $0.1185^{+0.0023}_{-0.0024}$ & $0.1189$ & $0.1195^{+0.0028}_{-0.0028}$ & $0.1189$\\[1ex]

{$100\theta_{MC} $} & $1.04089^{+0.00061}_{-0.00061}$ & $1.04092$&$1.04079^{+0.00061}_{-0.00062}$ &1.04069\\[1ex]

{$\tau           $} & $0.078^{+0.031}_{-0.028}   $ &  $0.075$ &$0.076^{+0.032}_{-0.032}   $ & $0.077$\\[1ex]
$N_*                       $ & $57.59^{+0.14}_{-0.14}     $ & $57.60$ & $- $ & $-$ \\[1ex]
$T_{RH}$\footnote{Here, $T_{RH}$ is expressed in $m^{-1}$}                    & $\left(\,3.15^{+0.13}_{-0.15}\,\right)\cdot 10^{31}$ & $3.15\cdot 10^{31}$& $-  $ & $-$ \\[1ex]
\hline
$\chi_{min}^2$ &\multicolumn{2}{|c}{$13004.946$} & \multicolumn{2} {|c|}{$13001.548$} \\[1ex]
${\rm{AIC}}$ &\multicolumn{2}{|c}{$13016.946$} & \multicolumn{2} {|c|}{$13015.548$} \\[1ex]
\hline
\end{tabular}
\caption{ Constraints on cosmological
parameters for the q-inflationary model compared with $\Lambda CDM+r$ model, obtained using
Planck data 2018 TT+TE+EE+lowTEB.}
\label{table:data}
\end{center}
\end{table}
\end{savenotes}
\section{CLOSING REMARKS}
 Despite the fact that the obtained results may allow the q-inflation to be a plausible candidate for describing the early universe, the current model is just a toy model and much enhancement has yet to be done.  In particular, the reheating mechanism must be further developed. As one may notice, the current model, similarly to the power-law inflation, has an exit problem, i.e. inflation does not end via slow roll violation, i.e. $\epsilon=1-q=const$. Nevertheless, as has been stated by Lucchin et al. $\cite{53}$ for a power-law inflation, one can assume that at a particular time $\tilde{t}_{_{RH}}=\tilde{t}_{end}$, the model no longer holds in such a way that a rapid reheating process takes place. Note that, in the current study, we have neglected the term ${m_\zeta^2}/{2H^2}$ appearing in Eq. \ref{recasted expanding kg2}, for the sake of simplicity. Nevertheless, despite its negligible effect one may take it into account. Doing so may lead to slight differences in the values of $q$, $r$, $N_*$ as well as $T_{RH}$.\\\\
Within the context of the q-inflationary scenario, ``although somehow hypothetical", one may emphasize that ``spacetime quanta"  may be relevant to such an inflation, inasmuch as the cosmological constant, i.e., the hyperboloid curvature radius triggering the inflation is carried by the quanta. If spacetime is really quantized then the present work can be seen as a little
step towards understanding (i) its quantum nature and (ii) why does time move forward. In this paper, we considered only quanta with ``frequency" $\emph{x}_0>0$, covering the hyperboloid's upper sheet but quanta that cover the lower sheet, i.e., $\emph{x}_0<0$ must also be taken into account, this is similar to K-G particle with negative energy $\omega<0$ that is considered as anti-particle having a positive energy. That is to say, if a parallel universe where time goes backward does exist then particles with negative energy may ``live" there, but they may be seen as anti-particles in our universe.\\
\section{ACKNOWLEDGMENTS}
I would like to thank Antony Lewis for kindly providing the numerical codes CosmoMC and CAMB. I am also grateful to Sukannya Bhattacharya and Savvas Nesseris for valuable instructions about numerical issues regarding CosmoMC, and I am particularly indebted to an anonymous reviewer for insightful comments.\\\\




\begin{thebibliography}{99}
\bibitem[1]{1}
S. W. Hawking, Nature (London) \textbf{248}, 30 (1974); Commun. Math. Phys. \textbf{43}, 199 (1975).
\bibitem[2]{2}
S. Weinfurtner. Quantum simulation of black-hole radiation. Nature, \textbf{569}, 634-635 (2019)
\bibitem[3]{3}
J. R. M. de Nova, K. Golubkov, V. I. Kolobov, and J. Steinhauer. Observation of thermal Hawking radiation and its temperature in an analogue black hole. Nature, \textbf{569}, 7758 (2019).
\bibitem[4]{4}
J.  Drori, Y. Rosenberg, D. Bermudez, Y. Silberberg, and U. Leonhardt, Phys. Rev. Lett. \textbf{122}, 010404 (2019)
\bibitem[5]{5}
P. Kraus and F. Wilczek, Nucl. Phys. B \textbf{433}, 403 (1995).
\bibitem[6]{6}
M.K. Parikh and F. Wilczek, Phys. Rev. Lett. \textbf{85}, 5042 (2000).
\bibitem[7]{7}
K. Srinivasan and T. Padmanabhan, Phys. Rev. D \textbf{60}, 024007 (1999).
\bibitem[8]{8}
J. Hu, L. Feng, Z. Zhang and C. Chin, Quantum simulation of Unruh radiation, Nat. Phys. (2019).
\bibitem[9]{9}
F. Scardigli, M. Blasone, G. Luciano and R. Casadio, Modified Unruh effect from generalized uncertainty principle, Eur. Phys. J. C \textbf{78}, 728 (2018).
\bibitem[10]{10}
J.A. Rosabal, New perspective on the Unruh effect, Phys. Rev. D \textbf{98}, 056015 (2018).
\bibitem[11]{11}
W. G. Unruh, Phys. Rev. D \textbf{14}, 870 (1976).
\bibitem[12]{12}
G.W. Gibbons and S.W. Hawking, Phys. Rev. D \textbf{15}, 2738 (1977).
\bibitem[13]{12x}
D. Chen,  H. Wu, H. Yang and S. Yang, Int. J. Mod. Phys. A \textbf{29}, 1430054 (2014).
\bibitem[14]{13}
T. Damour and R. Ruffini, Phys. Rev. D \textbf{20}, 239 (1976).
\bibitem[15]{14}
S. Sannan, Gen. Relativ. Gravit. \textbf{20}, 239 (1988).
\bibitem[16]{15}
P. Kraus and F. Wilczek, Nucl. Phys. B \textbf{437}, 231 (1995).
\bibitem[17]{16}
M. K. Parikh and F. Wilczek, Phys. Rev. Lett. \textbf{85}, 5042 (2000).
\bibitem[18]{17}
K. Srinivasan and T. Padmanabhan, Phys. Rev. D \textbf{60}, 024007 (1999).
\bibitem[19]{18}
M. Angheben, M. Nadalini, L. Vanzo and S. Zerbini, JHEP \textbf{05}, 014 (2005).
\bibitem[20]{19}
R. Kerner and R.B. Mann, Phys. Rev. D \textbf{73}, 104010 (2006).
\bibitem[21]{20}
A. H. Guth, Phys. Rev. D, \textbf{23}, 347 (1981).
\bibitem[22]{21}
A. H. Guth and So-Young Pi, Phys. Rev. Lett., \textbf{49}, 111 (1982).
\bibitem[23]{22}
 C. Armendariz-Picon, T. Damour, and V. F. Mukhanov, Phys. Lett., B \textbf{458}, 209 (1999).
\bibitem[24]{23}
A. D. Linde, Phys. Lett., B \textbf{108}, 389 (1982).
\bibitem[25]{24}
W. H. Kinney, Eternal Inflation and the Refined Swampland Conjecture, Phys. Rev. Lett. \textbf{122} (2019).
\bibitem[26]{25}
A. de la Fuente, P. Saraswat, and R. Sundrum, Natural Inflation and Quantum Gravity, Phys. Rev. Lett. \textbf{114}. 15, 151303 (2015).
\bibitem[27]{26}
E.O. Kahya, B. Pourhasssan and S. Uraz, Phys. Rev. D \textbf{92},103511 (2015).
\bibitem[28]{27}
 Z. Yi and Y. Gong, On the constant-roll inflation, JCAP \textbf{1803}, 052 (2018).
\bibitem[28]{28}
H. Motohashi and A.A. Starobinsky, JCAP \textbf{11}, 025 (2019).
\bibitem[30]{29}
 A. Mohammadi, K. Saaidi and T. Golanbari, Tachyon constant-roll inflation, Phys.Rev. D \textbf{97}, 083006 (2018).
\bibitem[31]{30}
 G. W. Gibbons and S. W. Hawking, Cosmological Event Horizons, Thermodynamics, And Particle Creation, Phys. Rev. D \textbf{15}, 2738 (1977).
\bibitem[32]{31}
J. R. Gott, Nature \textbf{295}, 304 (1982).
\bibitem[33]{32}
S. Umarov, C. Tsallis and S. Steinberg, Milan J. Math. \textbf{76} (2008). 
\bibitem[34]{33}
K. Ourabah and M. Tribeche, Phys. Rev. E \textbf{89},  062130 (2014).
\bibitem[35]{34}
A. Plastino and M. C. Rocca, Chin, Phys. C \textbf{42}, 053102 (2018).
\bibitem[36]{35}
L.C. C\'eleri and V.I. Kiosses, Canonical field quantization in momentum space and the corresponding Unruh effect, https://arxiv.org /abs /1712 .05206 (2017).
\bibitem[37]{36}
L.C. C\'eleri and V. I Kiosses, Unruh effect as a result of quantization of spacetime, Phys. Lett. B \textbf{781}, 611 (2018).
\bibitem[38]{37}
A. Dasgupta, The real Wick rotations in quantum gravity, JHEP 0207, 062 (2002).
\bibitem[39]{38}
H. Erbin and V. Lahoche, Universality of Tunneling Particles in Hawking Radiation, Phys. Rev. D \textbf{98}, 104001 (2018).
\bibitem[40]{39}
K. Yagdjian and A. Galstian. Fundamental solutions for the Klein-Gordon equation in de Sitter spacetime. Comm. Math. Phys., \textbf{285} (1) (2009).
\bibitem[41]{40}
Akrami et al. (Planck Collaboration), Planck 2018 results.X. Constraints on inflation, arXiv: 1807.06211.
\bibitem[42]{41}
P. A. R. Ade et al. (Planck Collaboration), Astron. Astrophys. \textbf{594}, A20 (2016). 
\bibitem[43]{42}
P. A. R. Ade et al. (BICEP2 and Keck Array Collaborations), Phys. Rev. Lett. \textbf{116}, 031302 (2016).
\bibitem[44]{43}
P. A. R. Ade, et. al., Planck 2013 results. XXII. Constraints on inflation, arXiv:1303.5082.
\bibitem[45]{44}
R. Allahverdi, R. Brandenberger, F.-Y. Cyr-Racine and  A. Mazumdar, Ann. Rev. Nucl. Part. Sci., \textbf{60}, 27 (2010).
\bibitem[46]{45}
J. Martin and C. Ringeval, First CMB Constraints on the Inflationary Reheating Temperature, Phys. Rev. D \textbf{82}, 023511 (2010).
\bibitem[47]{46}
U. Seljak and M. Zaldarriaga, Astrophys. J. \textbf{469}, 437 (1996).
\bibitem[48]{47}
A. Lewis and A. Challinor, Phys. Rev. D \textbf{66} (2), 023531 (2002).
\bibitem[49]{48}
A. Lewis, A. Challinor and A. Lasenby, Astrophys. J. \textbf{538}, 473 (2000).
\bibitem[50]{49}
A. Lewis, S. Bridle, Phys. Rev. D. \textbf{66} (10), 103511 (2002).
\bibitem[51]{50}
Planck collaboration, N. Aghanim et al., Planck 2018
results. VI. Cosmological parameters, 1807.06209.
\bibitem[52]{51}
H. Akaike, IEEE T. Automat. Contr. \textbf{19}, 716 (1974).
\bibitem[53]{52}
P. A. R. Ade et al. (Planck), Astron. Astrophys. \textbf{571}, A22, 1303.5082 (2014).
\bibitem[54]{53}
 S. Lucchin, and S. Matarrese, Phys. Rev. D \textbf{32}, 1316 (1985).

\end{thebibliography}
\end{document}